\documentclass[prl,twocolumn,showpacs,superscriptaddress,floatfix]{revtex4}
\usepackage{graphicx,amssymb,amsmath}

\newcommand{\rmi}{\mathrm{i}}
\newcommand{\ket}[1]{|#1\rangle} 
 
\newcommand{\braket}[2]{\langle#1|#2\rangle} 
\def\nbN{\ensuremath{\mathrm{I\!N}}}

\begin{document}
\title{The Thermodynamical Limit of the Lipkin-Meshkov-Glick Model}

\author{Pedro Ribeiro}
\author{Julien Vidal}
\author{R\'emy Mosseri}
\affiliation{Laboratoire de Physique Th\'eorique de la Mati\`ere Condens\'ee, CNRS UMR 7600,
Universit\'e Pierre et Marie Curie, 4 Place Jussieu, 75252 Paris Cedex 05, France}

\begin{abstract}
A method based on the analyzis of the Majorana polynomial roots is introduced to compute the spectrum of the Lipkin-Meshkov-Glick model in the thermodynamical limit.  A rich structure made of four qualitatively different regions is revealed in the parameter space whereas the ground state study only distinguishes between two phases.  
\end{abstract}

\pacs{05.30.-d,21.60.Ev,03.65.Sq}

\maketitle
%
%
%
%

The Lipkin-Meshkov-Glick (LMG) model introduced in $1965$ to describe shape phase transition in nuclei \cite{Lipkin65} has, since then, been proposed to describe many systems ranging from interacting spin systems \cite{Botet83} to Bose-Einstein condensates \cite{Cirac98} or magnetic molecules such as ${\rm Mn}_{12}$ ace\-tate \cite{Garanin98}. 
This ubiquity is due to its mapping onto a single-particle evolving in a double-well potential \cite{Turbiner88, Ulyanov92} or onto an interacting two-level boson system. More recently, this model has also been used to investigate the relationship between entanglement and quantum phase transitions \cite{Vidal04_1,Latorre05_2,Dusuel04_3,Barthel06_2}.

The LMG model is known to be exactly solvable \cite{Pan99,Links03_1,Ortiz05}. However, getting the solution requires to solve Bethe-like equations, a task which, in the present context, is more costly than exact diagonalization. 
A complete description of the spectrum thus requires to develop alternative routes. Though the low-energy spectrum has been studied in details via different methods (variational \cite{Lipkin65}, bosonization \cite{Dusuel04_3,Dzhioev04}, coherent states \cite{Kuriyama03}), the richness of the full spectrum has been investigated only  lately by means of numerical diagonalizations \cite{Heiss05,Castanos06}. These latter studies suggest the existence of singular points in the density of states as well as a nontrivial level spacing distribution.

In this paper, we shed light on these issues by exactly computing  the spectrum of the LMG model. The proposed method relies on the determination of the Majorana polynomial roots associated to the eigenstates of the Hamiltonian. This polynomial is built within a coherent state formalism which is well-suited to such a system. 
Within this framework, the spectrum is encoded in a linear differential equation which is solved  in the thermodynamical limit. This allows us to exactly compute the density of states in the whole parameter range and to locate its singularities. 
Four distinct regions arise with qualitatively different properties. In particular, we find a  parameter regime for which the density of states has no thermodynamical limit.

%
%
%
%

The LMG model describes a set of $N$ spins $\frac{1}{2}$ mutually interacting through a $XY$-like Hamiltonian and coupled to an external transverse magnetic field $h$.
This Hamiltonian $H$ can thus be expressed in terms of the total spin operators 
$S_{\alpha}=\sum_{i=1}^N \sigma_{\alpha}^{i}/2$ where the $\sigma_{\alpha}$'s are the Pauli matrices:
%
%
\begin{equation}
\label{eq:hamiltonian}
H=-\frac{1}{N} \big(\gamma_x S_x^2 + \gamma_y S_y^2  \big) - h \: S_z.
\end{equation}
%
%
In the following, we only consider the maximum spin sector $s=N/2$ with $N$ even.
Given the symmetry of the spectrum of $H$, we focus on the parameter range $h \geqslant 0$;  $0 \leqslant |\gamma_y| \leqslant \gamma_x$. Note also that $\big[ H, {\bf S}^2 \big]=0$ and $\big[ H, e^{i \pi  (S_z - s)} \big]=0$ (spin-flip symmetry). Denoting by 
$\{|s,m\rangle\}$ the standard eigenbasis of $\big\{{\bf S}^2, S_z\big\}$, this latter symmetry implies that odd and even states decouple.
In the thermodynamical limit, both subspaces are isospectral so that we limit the following analyzis to the sector $m$ even for which one has exactly $(s+1)$ eigenstates.

In the spin coherent states basis \cite{Klauder85}, with non-normalized states $\ket{\alpha}=e^{\bar\alpha S_+}\ket{s,-s}$,  any state $\ket{\Psi}$ is represented by its Majorana polynomial \cite{Majorana32} defined as
%
%
\begin{eqnarray}
\label{eq:polynome}
 \Psi(\alpha)=\braket{\alpha}{\Psi} & = & \hspace{-2mm} \sum _{m=-s}^s  \sqrt{\frac{(2 s)!}{(s-m)! (m+s)!}} \braket{s,m}{\Psi}  \alpha^{m+s}, \nonumber \\ 						
 & = & C \prod_{k = 1}^{d} \left(  \alpha - \alpha_k  \right)   ,
\end{eqnarray}
%
%
where $d \leqslant 2s$ is its degree.

\begin{figure}[t]
  \centering
\includegraphics[width=\columnwidth]{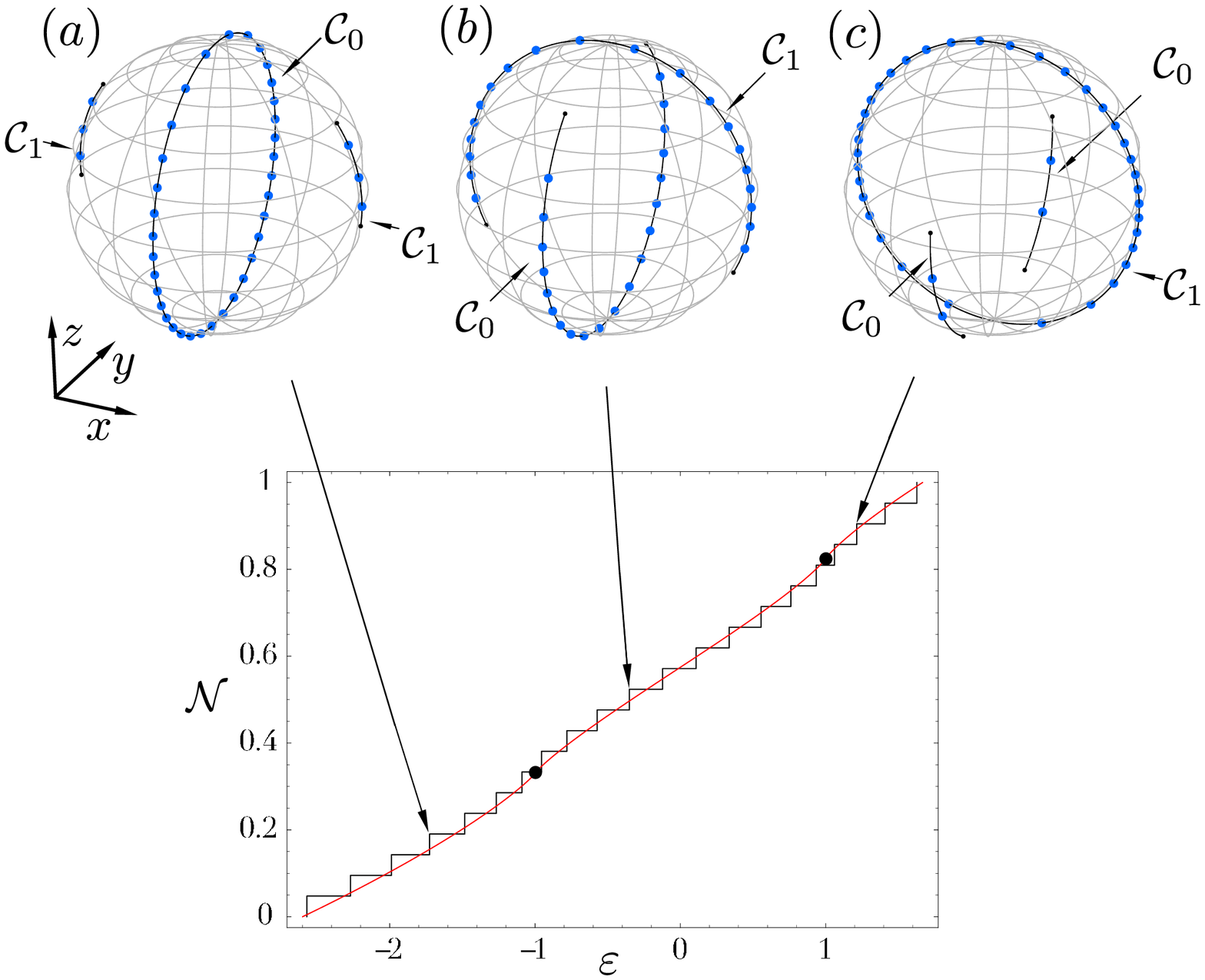}
\caption{ Upper part: representation of the poles of $G$ on the Majorana sphere (blue dots) for three  typical eigenstates computed for $h=1, \gamma_x=5, \gamma_y=-3$ and $s=20$ (zone III in Fig. \ref{fig:diagram}). Black lines correspond to the $G_0$ branch cuts $\mathcal{C}_0$ and $\mathcal{C}_1$. \\
Lower part: Numerical (black staircase curve $s=20$) versus analytical (red line $s=\infty$ ) integrated density of states. 
Black dots indicates the singularity of the density of states ${N}^{\rm III}_0(-h)$ and ${N}^{\rm III}_0(h)$ (Eqs. (\ref{eq:NII}) and (\ref{eq:NIII}) respectively) in  the thermodynamical limit.}
  \label{fig:Majsphere}
  \end{figure}

The standard representation of the spin operators ($S_\pm=S_x \pm \rmi S_y$) in the coherent states basis
%
%
\begin{eqnarray}
S_+&=& 2s \alpha-\alpha^2 \partial_\alpha, \\
S_-&=& \partial_\alpha , \\
S_z&=& -s+ \alpha \partial_\alpha , 
\end{eqnarray}
%
%
allows to map the Schr\"{o}dinger equation $H |\Psi \rangle =E |\Psi \rangle$ onto the following linear differential equation \cite{Turbiner88,Kurchan89}
%
%
\begin{equation}
\label{eq:Maj_Schrodinger}
\bigg[\frac{P_2(\alpha)}{(2s)^2} \partial^2_{\alpha}+ \frac{P_1(\alpha)}{2s} \partial_{\alpha} + P_0(\alpha ) \bigg] \Psi(\alpha) = \varepsilon \Psi(\alpha) ,
\end{equation}
where $\varepsilon=E/s$ and
%
%
\begin{eqnarray}
\label{eq:G_Schrodinger_polynomes}
&& \hspace{-8mm} P_0(\alpha ) =  \frac{1}{4s} \Big[ \alpha^2  (2s-1)(\gamma _y - \gamma _x )    -  \gamma _x - \gamma_y \Big] +h, \\
&&  \hspace{-8mm} P_1(\alpha) =  \alpha \bigg\{\frac{2s-1}{2s}  \Big[ \alpha^2(\gamma_x-\gamma_y)-\gamma_x-\gamma_y  \Big] -2h  \bigg\}, \\
&&  \hspace{-8mm} P_2(\alpha)= -\frac{1}{2} \left[\left(\alpha^2-1\right)^2 \gamma _x-\left(\alpha^2+1\right)^2 \gamma _y \right] . 
\end{eqnarray}
%
%
Except for trivial values of the parameters, the degree of $\Psi$ for an eigenstate of $H$ in the sector we considered is $d=2s$.
At this step, the spectrum could be analyzed by mapping Eq. (\ref{eq:Maj_Schrodinger}) onto a 
Schr\"{o}dinger equation describing a particle in an effective one-dimensional potential 
\cite{Turbiner88,Ulyanov92}. Then, a semi-classical treatment would, in principle, allow one to obtain the density of states in the thermodynamical limit, as shown in Ref. \cite{Garanin98} for the low-energy spectrum in the region $\gamma_y=0, \gamma_x >0$.
Unfortunately, for arbitrary values of the parameters, the effective potential becomes tricky \cite{Ulyanov92} and such an approach is therefore difficult to follow.  
Here, we propose an alternative route by first converting the linear second-order differential equation for $\Psi$ (\ref{eq:Maj_Schrodinger}) into a first-order differential equation for its loga rithmic derivative. More precisely, the function $G$ defined as
%
%
\begin{equation}
\label{G_definition}
G(\alpha) = \frac{1}{2s} \partial_\alpha \log \Psi(\alpha) = \frac{1}{2s}  \sum_{k=1}^{2s} \frac{1}{\alpha-\alpha_k}, 
\end{equation}
%
%
satisfies the following Riccati-like equation
%
%
\begin{equation}
\label{eq:G_Schrodinger}
P_2(\alpha) \left[ \frac{G'(\alpha)}{2 s} + G^2(\alpha) \right] + P_1(\alpha ) G(\alpha) +  P_0(\alpha) = \varepsilon.
\end{equation}
%
%
The density of states is then given by analyzing the poles of $G$, i.e., the roots of the Majorana polynomial $\Psi$. Indeed, the cornerstone of this study is that, for this model, the $\alpha_k$'s are spread over two curves $\mathcal{C}_0$ and $\mathcal{C}_1$ in the complex plane which depend on the energy. In addition, the $n$-th excited state of $H$ has $2n$ poles on $\mathcal{C}_1$ and $2(s-n)$ on $\mathcal{C}_0$ (thus defining  both curves). This remarkable property stems from the oscillation theorem which indexes the excited states for a particle in the effective one-dimensional potential (discussed above) by the number of wavefunction nodes. 
To illustrate this repartition of the poles which is likely related to the integrability of the model \cite{Leboeuf90}, we display in Fig. \ref{fig:Majsphere} several typical states in the Majorana sphere representation \cite{Majorana32}. This representation generalizes the celebrated Bloch sphere used for spin $\frac{1}{2}$ states and proceeds as follows.  For a given polynomial with $d$ roots, we first complement it with $(2s-d)$ roots at infinity in the complex plane. Next, the resulting set of $2s$ points is sent onto the unit sphere by an inverse stereographic map.  For instance, within this mapping, the basis state $|s,m\rangle$ is represented by $(s-m)$ points on one pole and $(s+m)$ points on the opposite pole. 
%
%
%
%

The location of the poles of $G$  explained above provides a straightforward relation between the normalized integrated density of states $\mathcal{N} \in[0,1]$ and the number of poles lying in 
$\mathcal{C}_1$. One indeed simply has
%
%
\begin{equation}
\label{eq:int_of_G}
\mathcal{N}(\varepsilon)=\frac{n+1}{s+1}=\frac{1}{s+1}\bigg[\frac{s}{2 \rmi \pi} \oint_{\widetilde{\mathcal{C}}_1} G(\alpha)  \:  {\rm d} \alpha +1\bigg],
\end{equation}
%
%
where $\widetilde{\mathcal{C}}_1$ is a contour that surrounds $\mathcal{C}_1$ and oriented such that $\mathcal{N}\geq 0$. In this equation, $G$ is built from the $n$-th excited state and the dependence of $\mathcal{N}$ with $\varepsilon$ is given via Eq. (\ref{eq:G_Schrodinger}).
Unfortunately, one cannot solve Eq. (\ref{eq:G_Schrodinger}) exactly at finite $s$, which would give a complete explicit solution to our problem. However, one can easily solve it perturbatively in $1/s$. 

Therefore, let us assume that $G$, and $\varepsilon$, can be expanded in the following form
%
%
\begin{equation}
\label{eq:one_over_s_development}
G = \sum_{i \in \nbN} \frac{G_i}{s^i}, \quad  \varepsilon = \sum_{i \in \nbN} \frac{\varepsilon_i}{s^i}.
\end{equation}
%
%
At leading order $s^0$, Eq. (\ref{eq:G_Schrodinger}) becomes a second-order polynomial equation for  $G_0$ whose solutions simply read
%
%
\begin{equation}
G^\pm_0 (\alpha) =  \frac{\alpha \big[\alpha^2(\gamma_y-\gamma_x)+\gamma_x+\gamma_y +2h\big]  \pm \sqrt{2 Q(\alpha)}}{2 P_2(\alpha)} ,
\end{equation}
%
%
where 
%
%
\begin{equation}
Q(\alpha) = (\gamma_y-\gamma_x)(h+\varepsilon_0) \alpha^4+ 2 \big[h^2+\gamma_x \gamma_y+\varepsilon_0 (\gamma_x+\gamma_y)\big] \alpha^2 \nonumber
\end{equation}
%
%
\vspace{-8mm}
%
%
\begin{equation}
+(\gamma_x-\gamma_y)(h-\varepsilon_0).
\end{equation}
%
%
The four roots of $Q$ are the branch points of $G$ which define the limit of the curves $\mathcal{C}_0$ and $\mathcal{C}_1$. A close analyzis of these branch cuts, in the parameter space, then leads to the integrated density of states in the thermodynamical limit which reads
%
%
\begin{equation}
\label{eq:N_0}
\lim_{s\rightarrow \infty} \mathcal{N} (\varepsilon) = \mathcal{N}_0 (\varepsilon_0)=
\frac{1}{2 \rmi \pi}
\int_{\mathcal{C}_1} \big[ G^+_0 (\alpha)-G^-_0 (\alpha)\big]  \: {\rm d} \alpha.
\end{equation}
%
%
This quantity can be expressed in terms of lengthy expressions involving elliptic integrals. We obtained these expressions in the whole parameter space and will give them explicitely in a forthcoming publication \cite{Ribeiro07_2}.
In the following, we only discuss qualitatively the various regions that must be distinguished, and  analyze them by means of the density of states  $\rho_0(\epsilon_0)=\partial_{\varepsilon_0}\mathcal{N}_0(\varepsilon_0)$.

\begin{figure}[t]
  \centering
\includegraphics[width=\columnwidth]{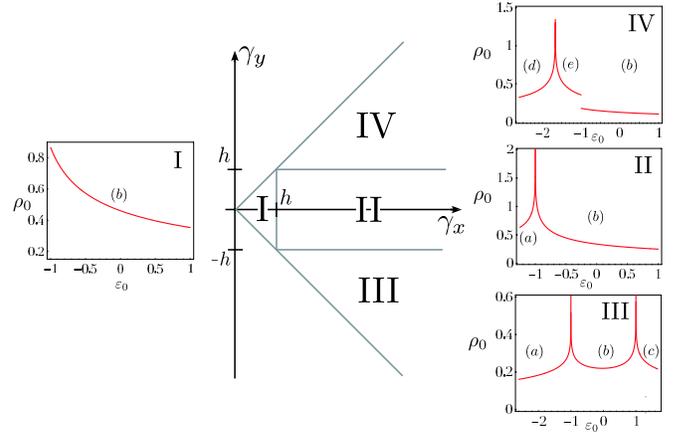}
  \caption{Phase diagram in the $\gamma_x,\gamma_y$ plane at fixed $h>0$ and typical density of states for $(\gamma_x,\gamma_y,h)$ equal to I: $(1/2,1/3,1)$,  II: $(2,1/2,1)$, III: $(5,-3,1)$,  IV: $(5,3,1)$.}
  \label{fig:diagram}
\end{figure}

But, first of all, let us remind the reader that if one only considers the properties of the ground state, which define the zero-temperature phase diagram, only two phases must be distinguished. For $h>\gamma_x$ (symmetric phase), the ground state is unique and $\lim_{s\rightarrow \infty} \langle S_z \rangle/s=1$ whereas for $h<\gamma_x$ (broken phase), the ground state is two-fold degenerate and $\lim_{s\rightarrow \infty} \langle S_z \rangle/s=h/\gamma_x$. The quantum phase transition at $h=\gamma_x$ is second-order and characterized by mean-field critical exponents \cite{Botet83} and nontrivial finite-size scaling behavior \cite{Dusuel04_3,Leyvraz05}. 
An important result of our study is that, when considering the full spectrum, four different zones arise instead of two. These regions, described below, are characterized by different singular behaviors of the density of states 
(see Fig. \ref{fig:diagram}) as already noticed in a numerical study of the special case $\gamma_x=-\gamma_y$ \cite{Heiss05}.

$\bullet $ Zone I: $  |\gamma_y | < \gamma_x <  h $. In this sector, the density of states $\rho_0$ is a smooth function of $-h \leqslant \varepsilon_0 \leqslant h$ as can be seen in Fig. \ref{fig:diagram}. The distribution of Majorana polynomial roots for the eigenstates is similar to that displayed in Fig. \ref{fig:Majsphere}$(b)$. In the complex plane, $\mathcal{C}_0$ and $\mathcal{C}_1$ lie in the imaginary and real axes  respectively.

$\bullet $ Zone II: $ \left|\gamma_y \right| < h < \gamma_x $. In this region, two distinct branches must be distinguished:

$-$ II$(a)$: $- \frac{h^2 + \gamma_x^2}{2 \gamma_x}  \leqslant \varepsilon_0 \leqslant -h$.
 $\mathcal{C}_0$ coincides with the whole imaginary axis while  $\mathcal{C}_1$ is made of two disconnected segments in the real axis as depicted in Fig.\ref{fig:Majsphere}$(a)$;

$-$ II$(b)$: $-h \leqslant \varepsilon_0 \leqslant h$. $\mathcal{C}_0$ and  $\mathcal{C}_1$ are the same as in I.

These two branches of the density of states diverge at  $\varepsilon_0=-h$ where the elliptic integrals involved in the expression of $\mathcal{N}_0$ can be recasted in the simple following form
%
%
\begin{eqnarray}
\label{eq:NII}
\mathcal{N}^{\rm II}_0(-h)  &=& 1- \frac{2}{\pi  \sqrt{\gamma _x \gamma _y}}  \times\\
&&
\bigg[
A_h^+ \tan^{-1} \frac{A_h^+}{B_h^0}-A_h^- \tan^{-1} \frac{A_h^-}{B_h^+}\bigg] ,\nonumber
 \end{eqnarray}
%
%
with
%
%
\begin{eqnarray}
\label{eq:NII}
A_h^{\pm}&=&h \pm \sqrt{\gamma _x \gamma _y},\\
B_h^{0}&=&\sqrt{h \gamma _x} + \sqrt{h \gamma _y}+\sqrt{(\gamma _x-h)(h-\gamma_y)}, \quad \\
B_h^{\pm} &=& \pm( \sqrt{h \gamma _x} - \sqrt{h \gamma _y})+\sqrt{(\gamma _x-h)(h-\gamma_y)}. \quad
 \end{eqnarray}
%
%

%
%
$\bullet$ Zone III:  $  h < - \gamma_y  < \gamma_x  $. In this zone, there are three different branches:

$-$ III$(a)$: $ - \frac{h^2 + \gamma_x^2}{2 \gamma_x} \leqslant \varepsilon_0 \leqslant -h $. $\mathcal{C}_0$ and  $\mathcal{C}_1$ are the same as in II$(a)$;

$-$ III$(b)$: $-h \leqslant \varepsilon_0 \leqslant h$. $\mathcal{C}_0$ and  $\mathcal{C}_1$ are the same as in I;

$-$ III$(c)$: $ h \leqslant \varepsilon_0 \leqslant - \frac{h^2 + \gamma_y^2}{2 \gamma_y} $.  $\mathcal{C}_0$ is made of two disconnected segments in the imaginary axis while $\mathcal{C}_1$ coincides with the whole real axis as depicted on the Majorana sphere  in Fig.\ref{fig:Majsphere}$(c)$.

In this zone, the density of states has two singularities at $\varepsilon_0=\pm h$. Their position in the spectrum is given by $\mathcal{N}^{\rm III}_0(-h)=\mathcal{N}^{\rm II}_0(-h)$ [see Eq. (\ref{eq:NII})] and
%
%
\begin{equation}
\label{eq:NIII}
\mathcal{N}^{\rm III}_0(h)  = \frac{2}{\pi  \sqrt{\gamma _x \gamma _y}} 
\bigg[
A_h^+\tan^{-1} \frac{A_h^+}{B_{-h}^0}-A_h^- \tan^{-1} \frac{A_h^-}{B_{-h}^-}\bigg] .
 \end{equation}
%
%
For $\gamma_x=-\gamma_y$, the density of states is symmetric with respect to $\varepsilon_0=0$, and the above expression gives the exact location, in the thermodynamical limit,  of the so-called exceptional point observed in Ref.  \cite{Heiss05}. 

%
%
%
%

$\bullet$ Zone IV: $ h  <  \gamma_y  < \gamma_x $. This part of the phase diagram is the most complex one. The density of states is, as in zone III, made of three different branches:

$-$ IV$(d)$: $- \frac{h^2 + \gamma_x^2}{2 \gamma_x} \leqslant \varepsilon_0 \leqslant - \frac{h^2 + \gamma_y^2}{2 \gamma_y}$;

$-$ IV$(e)$: $ - \frac{h^2 + \gamma_y^2}{2 \gamma_y} \leqslant \varepsilon_0 \leqslant -h$;

$-$ IV$(b)$: $-h \leqslant \varepsilon_0 \leqslant h$.

However, the structure of  the $\mathcal{C}_1$ curve, in each case, is complex but $\mathcal{C}_0$ remains simple so that the integral in Eq. (\ref{eq:N_0}) can still be computed and reveals two special points. The first one occurs at $\varepsilon_0= - \frac{h^2 + \gamma_y^2}{2 \gamma_y}$ for which
%
%
\begin{eqnarray}
\label{eq:NIVa}
\mathcal{N}^{\rm IV}_0 \bigg( - \tfrac{h^2 + \gamma_y^2}{2 \gamma_y} \bigg) &=& 1-\frac{1}{\pi  \sqrt{\gamma _x \gamma _y}}  \times\\
&& \hspace{-2mm}
\bigg[
A^+\tan^{-1} C_h-A^- \tan^{-1} C_{-h}\bigg] ,\nonumber
\end{eqnarray}
%
%
with
%
%
\begin{equation}
\label{eq:NIVb}
C_h= \frac{h \sqrt{\gamma_x}+\gamma _y^{3/2}}
{\sqrt{(\gamma _x -\gamma _y)(\gamma_y^2-h^2)}}.
 \end{equation}
%
%
There, the density of states diverges as in zone II and III.  The second one arises at $\varepsilon_0=-h$ where
%
%
\begin{equation}
\label{eq:NIVb}
\mathcal{N}^{\rm IV}_0 (-h)  =1 - \frac{h}{\sqrt{\gamma _x\gamma _y}},
 \end{equation}
%
%
but, at this energy, {\em the density of states is discontinuous in the thermodynamical limit} as can be seen in Fig. \ref{fig:diagram}. 
%
%
\begin{figure}[t]
  \centering
\includegraphics[width=\columnwidth]{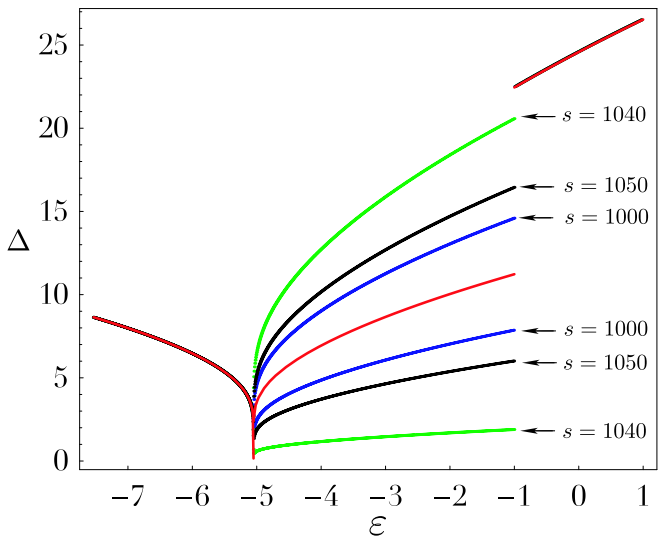}
  \caption{Gap between two consecutive levels as a function of the energy in region IV for 
  $\gamma_x=15$, $\gamma_y=10$ and $h=1$. In the central region, the red line is the average gap $\Delta_0$ in the thermodynamical limit. The branches below and above this line are respectively $\Delta^{(-)}$ and $\Delta^{(+)}$ (see text).}
  \label{fig:gap}
\end{figure}
%
%

We have confirmed this anomalous behavior numerically and observed an even more surprising result. 
Indeed, the energy difference between two consecutive levels 
$\Delta^{(i)}=E^{(i+1)}-E^{(i)}  (i=1,...,s)$ is normally given by  $\Delta_0(\varepsilon_0)=1/\rho_0(\varepsilon_0)$ in the thermodynamic limit. 
However, in region {\rm IV}$(e)$, numerical results (at fnite $s$) show that it is not the case.
Instead, $\Delta^{(i)}$ spreads over two branches $(+)$ and $(-)$, depending on the parity of the $i$. In addition, these branches oscillate without converging when $s$ increases as can be seen in Fig \ref{fig:gap}. 
In this case, the gap we computed , in the thermodynamical limit, is the average gap, namely 
$\Delta_0(\varepsilon_0)=\frac{1}{2}\big[ \Delta^{(+)}(\varepsilon_0)+\Delta^{(-)}(\varepsilon_0)\big]$.

To understand physically this unusual phenomenon, we have analyzed the classical trajectories in this region and have observed that,  there are two possible classical trajectories  \cite{Ribeiro07_2}. Using the results described in Ref. \cite{Kurchan89}, we computed analytically  the expectation values of several observables (such as the magnetization) as a function of energy. We found that these values also depend on the parity of the level considered but, contrary to the gap, the two branches, converge in the thermodynamical limit.

Finally, note that for $h=0$, the LMG model in this region coincides with the quantum asymmetric rotor model  \cite{King47}. 

In conclusion, we would like to emphasize that the present approach can be extended to other similar models with higher-order interaction terms where the mapping onto a particle in a one-dimensional potential fails.  Further, one can go beyond the thermodynamical limit and extract the finite-$s$ corrections which could be crucial for some observables \cite{Ribeiro07_2}.
 
\acknowledgments

We are grateful to  C. Aslangul, S. Dusuel, N. Gromov, J.-M. Maillard  and  P. Vieira for fruitful and stimulating discussions.
PR was partially supported by FCT and EU FEDER through POCTI and POCI, namely via QuantLog POCI/MAT/55796/2004 Project of CLC-DM-IST, SQIG-IT and grant SFRH/BD/16182/2004/2ZB5.


\end{document}